\documentclass[journal=acsodf,manuscript=article]{achemso}
\usepackage[version=3]{mhchem} 
\usepackage{hyperref}
\usepackage{pdfpages}

\author{Chris Yannic Bohlemann}
\affiliation{Technische Universität Ilmenau, Fundamentals of Energy Materials
\& Institute for Micro- and Nanotechnology, Postbox 100 565, D-98693 Ilmenau}
\alsoaffiliation{Technische Universität Ilmenau, CZS Junior Research Group for substitution and recycling strategies for solar energy materials, Postbox 100 565,D-98693 Ilmenau}
\author{Pavithira Manoharan}
\affiliation{Technische Universität Ilmenau, Fundamentals of Energy Materials
\& Institute for Micro- and Nanotechnology, Postbox 100 565, D-98693 Ilmenau}
\alsoaffiliation{Technische Universität Ilmenau, CZS Junior Research Group for substitution and recycling strategies for solar energy materials, Postbox 100 565,D-98693 Ilmenau}
\author{Helene Reichel}
\author{Kai Daniel Hanke}
\author{Peter Kleinschmidt}
\author{Thomas Hannappel}
\affiliation{Technische Universität Ilmenau, Fundamentals of Energy Materials
\& Institute for Micro- and Nanotechnology, Postbox 100 565, D-98693 Ilmenau}
\author{Juliane Koch}
\email{j.koch@tu-ilmenau.de}
\affiliation{Technische Universität Ilmenau, Fundamentals of Energy Materials
\& Institute for Micro- and Nanotechnology, Postbox 100 565, D-98693 Ilmenau}
\alsoaffiliation{Technische Universität Ilmenau, CZS Junior Research Group for substitution and recycling strategies for solar energy materials, Postbox 100 565,D-98693 Ilmenau}

\title{Controlling catalyst agglomeration in high-density unordered III-V nanowire growth using Au colloid solutions}

\keywords{Nanowires, Au Colloid}

\begin{document}

\begin{tocentry}

\includegraphics[]{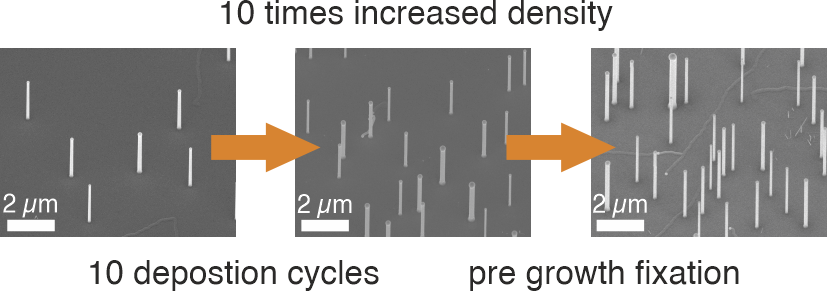}
\end{tocentry}

\begin{abstract}
III-V semiconductor nanowires (NWs) are a promising platform for optoelectronic and photoelectrochemical applications, where device performance strongly depends on NW density and spatial arrangement. While ordered arrays provide precise control, their fabrication requires complex and costly lithographic techniques. Unordered growth offers a scalable alternative but is limited by insufficient control over catalyst distribution and particle agglomeration.

Here, we investigate the density scaling of unordered III-V NW arrays using commercially available Au colloid solutions as catalysts for NW growth via vapor-liquid-solid growth mode. Repeated deposition cycles yield a near-linear increase in particle density, which is ultimately limited by non-linear agglomeration effects not captured by simple stochastic models.

To address this limitation, a previously established pre-anneal growth concept is transferred from patterned catalyst arrays to randomly deposited Au colloids, thereby suppressing thermally induced coalescence and stabilizing the catalyst distribution. This approach enables up to a tenfold increase in NW density while improving uniformity and vertical yield. The method is demonstrated for colloid diameters between 100 and 200 nm.

Overall, this work provides a scalable, lithography-free route toward high-density III-V NW ensembles and offers insight into the role of particle dynamics in colloid-based growth processes.
\end{abstract}

\section{Introduction}
Over the past decades, semiconductor nanowires (NWs) have emerged as a versatile platform for optoelectronic devices due to their quasi-one-dimensional geometry, which enables efficient charge carrier confinement, short transport paths, and pronounced surface interactions. These properties make NWs promising for applications in photodetectors \cite{Li.2019,Shoaib.2017,Barrigon.2019, Krogstrup.2013}, light-emitting devices \cite{Koester.2015, AzizarRahman.2025,Johar.2020}, chemical sensors \cite{Ahn.2009,AkbariSaatlu.2020,Cho.2020}, and photoelectrochemical systems\cite{Koch.2026,Hannappel.2024,Vanka.2018}.

Their high surface-to-volume ratio, combined with geometry-dependent optical effects such as light trapping and enhanced absorption,\cite{Xu.2015, Koch.2025a} further strengthens light-matter interaction and positions NW arrays as an efficient architecture for energy conversion structures.
Among available material systems, III-V semiconductor materials are particularly attractive due to their tunable band gaps, high charge carrier mobility, and controllable doping. \cite{Li.2020, Nagelein.2019, Jeddi.2023} In NW geometries, elastic strain can relax laterally, enabling the integration of lattice-mismatched materials with significantly reduced defect densities compared to planar thin films \cite{Glas.2006}. This opens a pathway for the monolithic integration of high-performance III-V materials on silicon substrates, supporting scalable and cost-effective device fabrication\cite{Steidl.2017, Krogstrup.2013b}.

However, device performance is highly sensitive to NW geometry and array configuration including NW density, uniformity, and spatial arrangement. Numerical studies indicate that optimal absorption is material dependent and typically achieved for III-V NW lengths exceeding 2 \textmu m and diameters ranging from 140 nm (InP) to 180 nm (GaAs) \cite{ Wu.2017, Wu.2017b, Hu.2012, Aberg.2016}. In addition, the ratio between inter-wire distance, which is also known as pitch, and NW diameter is a critical parameter, with optimal pitch-to-diameter ratios of 0.4 to 0.5 for GaAs,\cite{Guo.2011, Wu.2017b, Hu.2012, Aberg.2016} and lower ratios of 0.2 to 0.4 for In-based systems.\cite{Wu.2017,Otnes.2018, Treu.2019}
Because NW density scales inversely with the square of the array pitch, these constraints correspond to target densities of approximately 2$\times$10\textsuperscript{8} NW/cm\textsuperscript{2} for InP NW arrays and up to  8$\times$10\textsuperscript{8} NW/cm\textsuperscript{2} for GaAs NW arrays. Achieving such densities is therefore essential for maximizing optical absorption. 

Realizing these geometrically optimized NW arrays requires precise and scalable fabrication strategies. In general, NWs can be fabricated using either top-down or bottom-up approaches \cite{Hobbs.2012}. Top-down methods rely on lithography and subsequent etching to define NW structures with high positional accuracy, \cite{Tintelott.2021} enabling horizontally or vertically aligned NWs architectures\cite{Arjmand.2022, Liu.2015}. However, these often suffer from surface roughness and etching-induced defects that can degrade optical and electronic performance \cite{Demontis.2021, Ghazali.2016}.  
In contrast, bottom-up approaches offer superior material quality and improved material efficiency  \cite{Arjmand.2022, Li.2020, Demontis.2021}. Among these, particle-assisted growth based on the vapor-liquid-solid (VLS) mechanism is widely employed\cite{Hobbs.2012,AkbariSaatlu.2020,Demontis.2021}, where metallic catalyst particles determine the NW diameter and nucleation site. 
Especially for GaAs NWs, Au-based catalysts play an active role in the growth process by collecting growth species and forming an Au–Ga-rich phase at the NW growth front, thereby promoting precursor incorporation and axial crystal growth \cite{Borgstrom.2004,Persson.2004,Dick.2005,Dick.2008}.
A key distinction within bottom-up fabrication lies between ordered arrays and unordered NW ensembles \cite{Koch.2025a,Demontis.2021}. 
Ordered arrays can for example be fabricated using electron beam lithography \cite{Wu.2002, Bauer.2010} or nano imprint lithography \cite{Pierret.2010, Munshi.2014}, provide precise control over position and geometry, enabling reproducible device characteristics and facilitating simplified modeling. However, they typically rely on complex and costly lithographic processing\cite{Koch.2025a, Arjmand.2022}. In contrast, unordered growth routes relying on depositing colloidal Au nanoparticles \cite{Messing.2009,Messing.2010} or thermally dewetting Au thin films,\cite{Xu.2012,Sui.2014} offer a simpler and more cost-effective fabrication route, but generally results in reduced control over NW uniformity and spatial distribution \cite{GarciaGil.2021}. 
Colloidal Au nanoparticles are particularly attractive for GaAs NW growth because they provide a simple and inexpensive fabrication route together with well-defined catalyst diameters and, consequently, good control over the initial NW diameter \cite{Messing.2009,Messing.2010}. However, reproducible control of catalyst density and spatial distribution remains challenging, since deposition, agglomeration, and thermally induced particle motion on GaAs can modify the initially deposited particle ensemble \cite{Messing.2010, Whiticar.2017}.
This often leads to lower achievable densities and limits the exploitation of collective optical effects \cite{Chen.2016, Jager.2014}.
Systematic studies on achievable NW densities in unordered particle-assisted growth, particularly for colloid diameters above 100 nm, remain scarce. For smaller particle diameters ( $\leq$ 100 nm), densities up to 1.1$\times$ 10\textsuperscript{10} NW/cm\textsuperscript{2} have been reported using sputter-based deposition, albeit with very small NW radii of 2.8 nm  \cite{Puglisi.2019}. In comparison, solution-based Au colloid deposition has achieved densities of  1.8$\times$10\textsuperscript{8} NW/cm\textsuperscript{2} for Si NWs (50 nm Au colloids)  \cite{Hochbaum.2005}, and 7.3$\times$10\textsuperscript{7} NW/cm\textsuperscript{2} for InGaAs NWs (30 nm Au colloids) \cite{Kim.2006b}.

However, these densities are typically associated with small NW diameters, which limit optical absorption. Scaling such approaches to larger diameters remains challenging due to constraints in particle concentration within solutions, particle agglomeration effects and difficulties in achieving sufficiently low pitch-to-diameter ratios. 
As a result, reaching sufficient high NW densities required for efficient optical absorption remains a major challenge in unordered particle-assisted growth, thereby limiting their applicability for scalable optoelectronic devices.\cite{Messing.2010}

In this work, we systematically investigate the limitations and optimization pathways for achieving high NW densities in unordered III-V NW ensembles grown from commercially available Au colloid solutions. We identify catalyst deposition as the dominant limiting factor governing density scaling across different colloid sizes. Furthermore, we show that the observed agglomeration behavior cannot be adequately described by simple analytical models, highlighting the need for more advanced modeling approaches.
We demonstrate that a pre-anneal growth step previously developed for patterned Au catalyst arrays \cite{Koch.2025a, Otnes.2016} is transferable to solution-deposited colloids and suppresses their mobility and agglomeration during subsequent high-temperature annealing. This results in a density increase of up to one order of magnitude across a wide range of NW diameters. 

\section{Results and Discussion}
\begin{figure}
    \centering
    \includegraphics[width=1\linewidth]{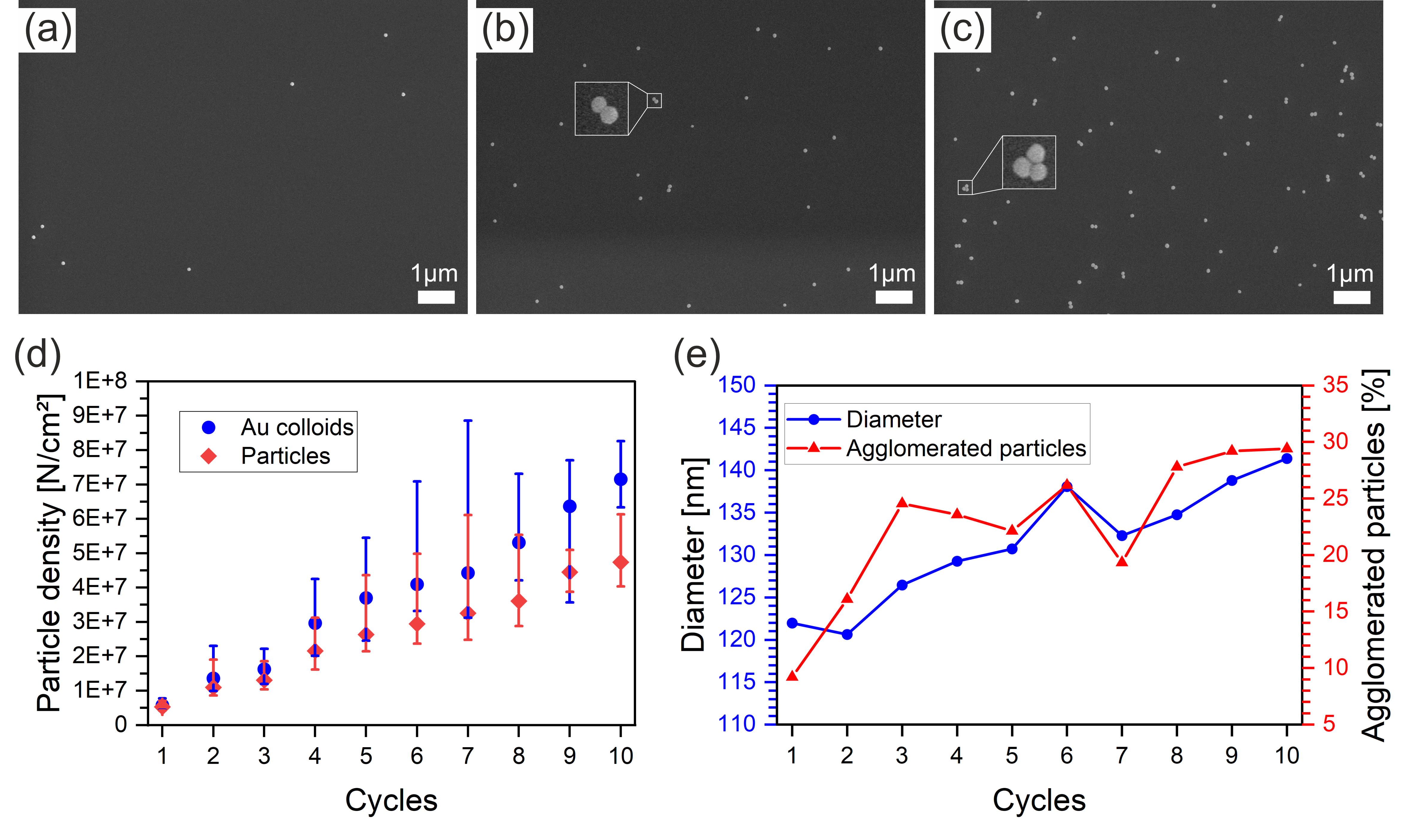}
    \caption{SEM images of Au colloid deposition after (a) 1, (b) 5, and (c) 10 successive deposition cycles using a 100 nm Au colloidal solution. The inset in (b) and (c) highlights a representative agglomerate at higher magnification. (d) Median particle densities of deposited Au colloids (blue) and resulting particles considering agglomeration (red). Error bars indicate the minimum and maximum values obtained from the evaluated dataset. (e) Average particle diameter (blue, left axis) and fraction of agglomerated particles (red, right axis) as a function of deposition cycle number}
    \label{fig:1}
\end{figure}

To investigate the influence of catalyst deposition on achievable NW densities, we systematically examine how the Au seed particle density evolves with repeated deposition cycles of Au colloid solution. As outlined, catalyst particle deposition is directly linked to the maximum attainable NW density \cite{Hochbaum.2005} and thus represents a fundamental limiting factor for high-density unordered NW growth.
A commercially available Au colloid solution was utilized, along with the deposition and fixation procedure, which is described in detail in the \nameref{Methods} Section, and was applied iteratively for up to ten cycles on multiple independent samples. 
The resulting catalyst-covered surfaces without further processing are shown in Fig. \ref{fig:1}. 
By scanning electron microscopy (SEM) image analysis it can be confirmed that the density of deposited Au colloids increase with the number of deposition cycles. Concurrently, significant agglomeration is observed, particularly after five and ten cycles (Fig. \ref{fig:1}(b, c)). These agglomerates predominantly consist of two to three primary colloids, as illustrated in the inset of Fig. \ref{fig:1} (b) and (c).

Quantitative analysis of the SEM images (Fig. \ref{fig:1}(d)) confirms a simultaneous increase in both colloid and particle densities with repeated deposition. 
To quantify the evolution of particle density, colloids and agglomerates were systematically identified and counted using a custom Python-based image analysis routine. Each SEM image covered an area of approximately 420 \textmu m\textsuperscript{2}, resulting in a total evaluated area of about 6000-12000 \textmu m\textsuperscript{2} per deposition cycle across multiple independently prepared samples.
However, the extracted data deviate significantly from a linear scaling behavior. As shown in Fig. \ref{fig:1}(a-c) and further quantified in Fig. \ref{fig:1} (e), both the average particle diameter and the fraction of agglomerated particles exhibit a pronounced non-linear dependence on the number of deposition cycles. In particular, the agglomeration fraction shows a rapid increase during the initial cycles, followed by a more gradual rise at higher cycle numbers, indicating that particle-particle interactions increasingly dominate the deposition process at higher surface coverage. This behavior ultimately limits the achievable particle density despite repeated deposition cycles.

The observed non-linear evolution of particle density and agglomeration points toward particle formation being governed by at least two competing mechanisms rather than independent deposition alone. The initial rapid increase in agglomeration suggests additional contributions beyond independent deposition, potentially arising from deposition-induced changes in the colloidal system. These may include modifications of substrate surface charging after the initial deposition cycle (e.g., due to HCl addition), temperature-dependent effects, concentration gradients building up inside the solution during waiting times, or increased clustering within the colloidal solution prior to subsequent deposition steps.\cite{Whitmer.2011,Roy.2018,Marc.2026,Ferrar.2015,Chu.2020} In contrast, the slower increase observed at later stages is more consistent with a process resembling a time dependent stochastic deposition.
As shown in Fig. \ref{fig:1} (e), the increase in average particle diameter directly correlates with the fraction of agglomerated particles. Notably, the average number of primary colloids per agglomerate remains nearly constant at approximately 2.4 - 2.6 colloids per agglomerated particle for most deposition cycles, with minor deviations for the first and final cycles. This further strengthens the assumption of a time-dependent stochastic deposition. Such stochastic deposition processes can be described using constant-rate event models, such as the Poisson distribution, which is applied in Supplementary Note 1 to model the observed agglomeration behavior.

While simplified Poisson-based models can provide a rough approximation of general trends and overall limitations, as shown in Supplementary Note 1, more refined and specific models are required to accurately capture detailed behaviors, such as the pronounced initial increase in agglomeration observed experimentally. This deviation highlights the importance of additional interaction mechanisms that are not included in such models.
Overall, these results demonstrate that particle-particle interactions in solution, surface effects induced by prior deposition, and process parameters such as potential temperature increase and timing play a critical role in determining the final particle distribution. The experimentally observed behavior therefore cannot be fully captured within the simplified modeling framework considered here, as this would require explicitly accounting for the interplay of particle-particle interactions, deposition-induced surface effects, and process-dependent parameters.

Despite these limitations, the results of this study clearly show that repeated deposition cycles of Au colloid solutions enable a substantial increase in both colloid and seed particle densities by approximately one order of magnitude prior to NW growth. The achievable density is currently limited by agglomeration effects and colloid solution concentration. Similar trends were observed for larger colloid diameters (150 nm and 200 nm) shown in supplementary note 2, indicating the generality of the approach.

As indicated by both the Poisson-based model and the experimental observations in Fig. \ref{fig:1}(b)-(d), the deposition of Au colloids from solution results in an inherently unordered and partially inhomogeneous distribution of seed particles. While strict size uniformity is not essential for all applications-particularly for photoabsorbers in photovoltaic (PV) and photoelectrochemical (PEC) systems, where broadband absorption can tolerate some variation in diameter, particle density and spatial distribution remain critical parameters for device performance.

\begin{figure}
    \centering
    \includegraphics[width=1\linewidth]{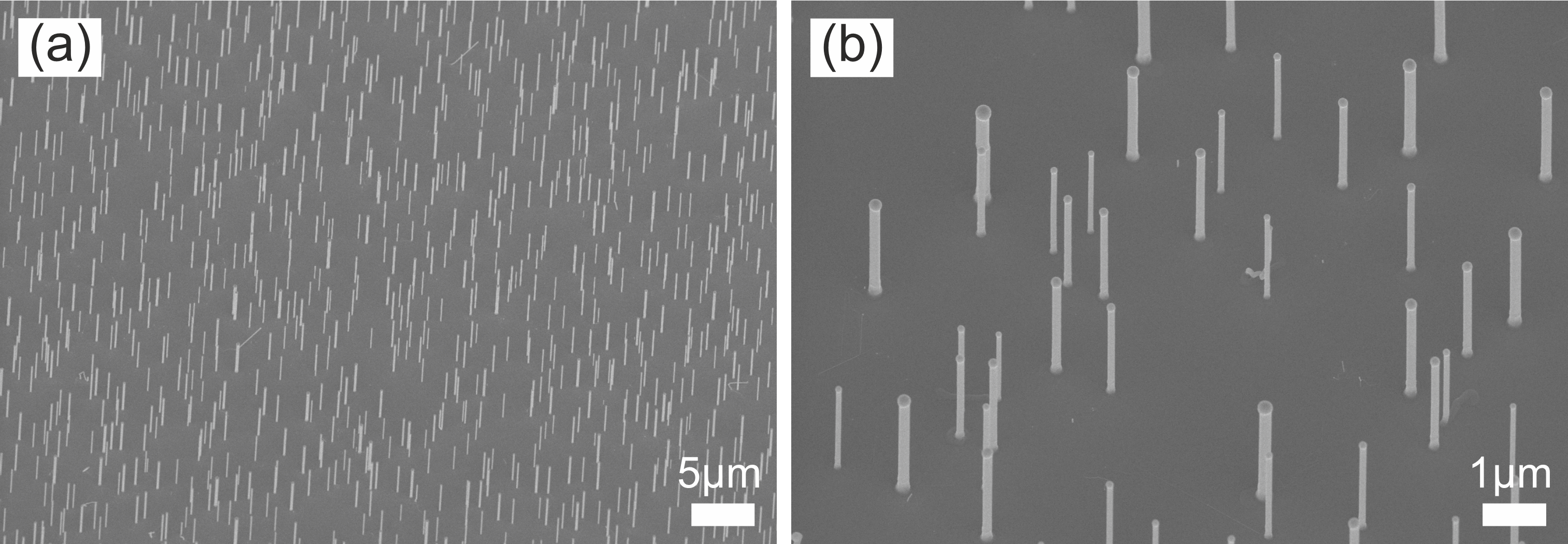}
    \caption{SEM images of GaAs nanowires (NWs) grown after 10 Au colloid deposition cycles. (a) Large-area image showing a high vertical yield of NWs. (b) Higher-magnification image revealing significant inhomogeneity in NW length and diameter. Images were acquired at a 30\textdegree~ tilt angle; the average NW length is approximately 3 \textmu m. }
    \label{fig:2}
\end{figure}

To evaluate the impact of these seed particle characteristics on NW growth, homoepitaxial GaAs NWs were grown on GaAs(111)B substrates using MOVPE via the VLS mechanism. Representative results are shown in Fig. \ref{fig:2}. As seen in Fig. \ref{fig:2}(a), a high vertical yield of NWs is achieved over large areas. However, significant inhomogeneity in both NW length and diameter is observed (Fig. \ref{fig:2}(b)), reflecting the non-uniform seed particle distribution.

A comparison of NW density with the corresponding Au particle density reveals a noticeable reduction during growth. The NW density for the sample shown in Fig. \ref{fig:2} is 3.74$\times$10\textsuperscript{7} NW/cm\textsuperscript{2}, corresponding to values expected for approximately 7-8 deposition cycles in Fig. \ref{fig:1}(d), despite 10 cycles being applied. This indicates an effective density reduction of roughly 20\% during the growth process. In addition, the average Au particle diameter at the NW top determined from the evaluation of more than 100 NWs, is approximately 160 nm, and therefore about 20 nm larger than the expected Au particle diameter for 10 deposition cycles based on Fig. \ref{fig:1}(e). These significantly larger Au particle at the NW top suggest that further agglomeration occurs during the initial stages of NW growth, leading to both increased Au particle size for the NWs and reduced NW density.
This behavior can be explained by the MOVPE growth procedure. Prior to NW growth at 400 \textdegree C, a high-temperature annealing step at 600 \textdegree C is applied for several minutes to form a eutectic alloy between the Au particles and the substrate and to remove surface oxides under an arsenic atmosphere. During this step, the mobility of Au particles is significantly increased, promoting particle coalescence. While this effect is negligible at low particle densities, higher densities increase the probability of particle-particle interactions, leading to additional agglomeration and further degradation of spatial uniformity.

To mitigate these effects, strategies for stabilizing the Au seed particles prior to high-temperature annealing are required. One effective approach is the introduction of a pre-anneal growth step.\cite{Koch.2025a} In this method, a short MOVPE growth step at lower temperatures (320 \textdegree C) is performed prior to annealing, enabling partial fixation of the particles. Previous studies have shown that this approach can significantly suppress particle mobility and enable homogeneous NW growth even at high densities \cite{Koch.2025a, Otnes.2016}.

\begin{figure}
    \centering
    \includegraphics[width=1\linewidth]{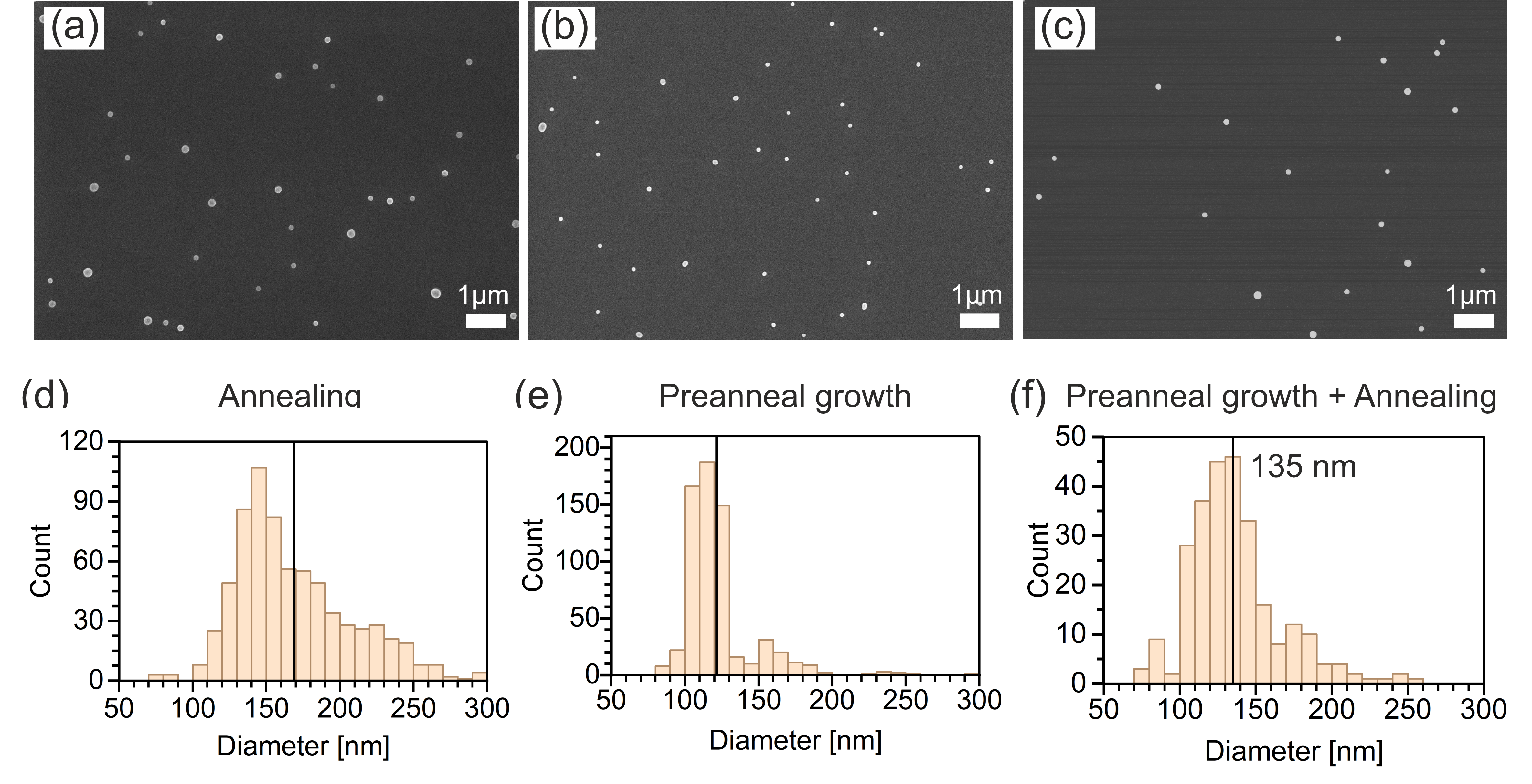}
    \caption{SEM images of Au seed particles and corresponding diameter histograms for different MOVPE process conditions: after 5 min annealing (a, d); after a 90 s pre-anneal growth step (b, e); and after a 90 s pre-anneal growth step followed by 5 min annealing (c, f). The pre-anneal growth step leads to a reduced average particle diameter and a narrower diameter distribution, indicating improved particle uniformity for samples with 10 deposition cycles. The black vertical lines in the histograms mark the average particle diameter. }
    \label{fig:3}
\end{figure}

\begin{figure}
    \centering
    \includegraphics[width=1\linewidth]{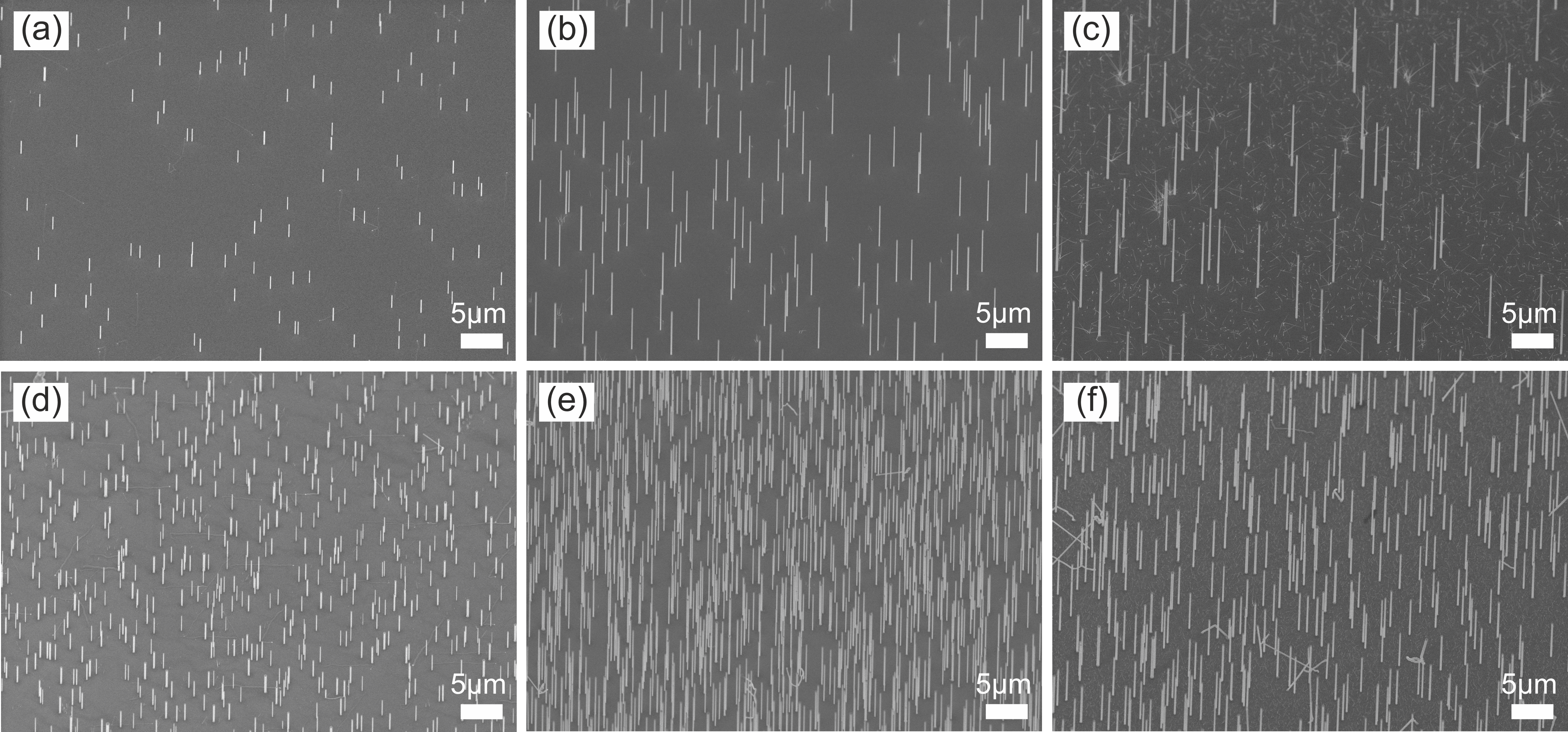}
    \caption{First row shows SEM images of the samples with GaAs NWs grown after 1 deposition cycle (a)-(c), while the second row shows samples with GaAs NW samples grown with 10 deposition cycles (d) and (e) as well as 6 deposition cycles (f). The seed particle diameters of the shown samples are: left column 100 nm (a),(d) with NW length of about 4 \textmu m; middle column 150 nm (b), (e) with NW length of 15 \textmu m and 10 \textmu m respectively and right column 200 nm (c),(f) with NW length of roughly 10 \textmu m and 8 \textmu m. SEM images were taken under a 30\textdegree ~tilt angle.}
    \label{fig:4}
\end{figure}

In this work, we adapt this pre-anneal growth strategy to unordered colloid-based deposition. The resulting particle behavior for applying only an annealing step at 600 \textdegree C for 5 minutes under constant As stabilization (a), a pre-anneal growth step at 320 \textdegree C for 90 s with both As and Ga precursors present  (b) and both pre-anneal growth and subsequent annealing step (c) followed by cooling to room temperature is shown in Fig. \ref{fig:3}. 
Building on the previously identified role of particle mobility during high-temperature processing, Fig. \ref{fig:3} illustrates the effect of annealing and pre-anneal growth steps on the Au seed particle distribution. After 5 min of high-temperature annealing (Fig. \ref{fig:3}(a, d)), a decrease in particle homogeneity is observed, accompanied by a larger average particle diameter within the range corresponding to the NWs shown in Fig. \ref{fig:2}. The resulting particle density of 3.69$\times$10\textsuperscript{7} 1/cm\textsuperscript{2} closely matches the NW density, confirming that particle coalescence during annealing directly limits the achievable NW density.
The introduction of a pre-anneal growth step prior to annealing leads to a distinctly different particle distribution. The time, III-V ratio and temperature of this pre-anneal growth step are dependent on the particle volume and process parameters to achieve sufficient fixation without lateral growth. For the samples shown in Fig. \ref{fig:3} (b) and (c), this time was set to 90 s, while later optimization of subsequent NW growth showed that 60 s pre-anneal growth time (Fig. \ref{fig:4} (a) and (d)) provides a higher proportion of vertical growth with sufficient fixation. As shown in Fig. \ref{fig:3}((b), (e)), the diameter histogram exhibits a narrow, near-Gaussian distribution, indicating significantly improved uniformity. The average particle diameter is reduced to 121 nm, which is substantially smaller than both the annealed case and the initial particle distribution (Fig. \ref{fig:1} (e)) for 10 deposition  cycles. This reduction can be attributed to partial particle fixation during the low-temperature growth step, which suppresses agglomeration and promotes spatial separation of particles.
However, this effect is not fully preserved during the subsequent high-temperature annealing. As shown in Fig. \ref{fig:3}(c, f), the average particle diameter increases again to approximately 140 nm, indicating partial re-agglomeration colloids which had been previously separated by pre-anneal growth. Nevertheless, the overall particle size distribution remains significantly more uniform compared to the case without pre-anneal growth. This demonstrates that the pre-anneal growth step effectively mitigates the effects of particle mobility while maintaining a suitable particle size for NW growth.

Combining these findings, the optimized process, consisting of repeated colloid deposition followed by a pre-anneal growth step and subsequent annealing, enables both high particle densities and improved uniformity. To evaluate the generality of this approach, NW growth was performed for three different Au colloid sizes (100 nm, 150 nm, and 200 nm), as shown in Fig. \ref{fig:4}.

For 100 nm colloids, the NW density increases from 4.65$\times$10\textsuperscript{6} NW/cm\textsuperscript{2} (Fig. \ref{fig:4}(a)) to 4.79$\times$10\textsuperscript{7} NW/cm\textsuperscript{2} (Fig. \ref{fig:4}(d)), corresponding to an approximately tenfold enhancement. For 150 nm colloids, a similar increase by a factor of about ten is observed, from 3.66$\times$10\textsuperscript{6} NW/cm\textsuperscript{2} (Fig. \ref{fig:4}(b)) to 3.52$\times$10\textsuperscript{7} NW/cm\textsuperscript{2}  (Fig. \ref{fig:4}(e)). For 200 nm colloids, where agglomeration is more pronounced, the number of deposition cycles was reduced to six. Nevertheless, a sixfold increase in density was achieved, from 2.46$\times$10\textsuperscript{6} NW/cm\textsuperscript{2} (Fig. \ref{fig:4}(c)) to 1.76$\times$10\textsuperscript{7} NW/cm\textsuperscript{2}  (Fig. \ref{fig:4}(f)), while maintaining a stable average diameter of approximately 255 nm. It is additionally noted that the density of NWs can locally vary throughout the NW ensemble due to the random distribution of the particle deposition.
Across all investigated particle sizes, the introduction of the pre-anneal growth step enables the formation of homogeneous, high-density, and vertically aligned NW ensembles with yields exceeding 90\%. These results demonstrate that the primary limitation of unordered NW growth using Au colloids, particle agglomeration, can be effectively mitigated by process engineering.
Overall, this study establishes a scalable and lithography-free approach for achieving high-density III-V NW ensembles. By combining controlled colloid deposition with tailored growth pre-treatment, NW densities approaching application-relevant regimes can be realized, providing a viable pathway toward more cost-effective integration of NW-based optoelectronic devices.

\section{Conclusions}
This study systematically investigated the possibilities and limitations of achieving high NW densities in unordered, colloid-based growth systems with a focus on scalable and cost-effective fabrication. It was demonstrated that repeated deposition of Au colloid solutions enables a near-linear increase in seed particle density. However, this approach is fundamentally limited by particle agglomeration during the particle deposition and the resulting loss of spatial uniformity.
The observed agglomeration behavior deviates from simple stochastic expectations and cannot be adequately described by a Poisson-based model, highlighting the need for more advanced computational approaches to capture the underlying particle-particle interactions and process-dependent effects.
Furthermore, it was shown that controlling particle mobility is essential for maintaining both high density and uniformity. Adapting the previously established pre-anneal stabilization concept to solution-deposited Au colloids effectively suppresses additional thermally induced agglomeration prior to NW growth. This strategy was successfully applied to Au colloids with diameters of 100 nm, 150 nm, and 200 nm, resulting in density increases of up to one order of magnitude while maintaining high vertical yield and improved size uniformity.
Overall, the presented approach provides a scalable, lithography-free pathway toward high-density III-V NW ensembles. By combining controlled colloid deposition with tailored process engineering, this method offers a promising alternative to conventional ordered array fabrication, enabling low-cost and high-throughput integration of bottom-up grown NWs for optoelectronic and photoelectrochemical applications.

\section{Methods} 
\label{Methods} 

\subsubsection{\textbf{Substrate Preparation}}

All samples were prepared on p-doped GaAs(111)B substrates (AXT) with a miscut angle of 0.1\textdegree~and a Zn doping concentration of N\textsubscript{A}(Zn) = 2.4$\times$10\textsuperscript{19} cm\textsuperscript{-3}. Prior to processing, substrates were cleaned by sequential immersion in acetone and isopropanol for 30 s each, followed by drying under a nitrogen flow after each step.

\subsubsection{\textbf{Au Colloid Deposition}}

Au colloid solutions with nominal particle diameters of 100 nm, 150 nm, and 200 nm (Sigma-Aldrich, citrate-buffer) were used. Prior to deposition, the solutions were homogenized by manual agitation followed by 10 min of ultrasonication to minimize pre-existing agglomerates.
A single deposition cycle was performed by drop-casting the colloid solution onto the substrate surface using a clean pipette. After a waiting period of approximately 20-40 s, a diluted HCl solution (4\% HCl) was applied to modify the surface polarization\cite{Joyce.2011, Woodruff.2007}. Following an additional waiting period of several minutes, the sample was dried under nitrogen flow.
The ratio between HCl solution and colloid solution was maintained at approximately 1:2. Care was taken to avoid overflow of the liquid from the substrate surface, as this was found to significantly affect particle density and reproducibility. Deposition cycles were repeated up to 10 times depending on the target particle density.

\subsubsection{\textbf{MOVPE Growth, Thermal Processing and Characterization}}

The MOVPE temperature sequence and precursor conditions were based on the process reported by Koch et al.,\cite{Koch.2025a} with particle-size-dependent pre-anneal durations and flow rates adapted specifically for the colloid-derived catalyst ensembles investigated here. NW growth and thermal treatments were performed using a MOVPE system (Aixtron AIX 200, horizontal reactor).
A pre-anneal growth step was introduced prior to high-temperature annealing, following the approach reported by Koch et al. \cite{Koch.2025a}. This step was carried out at 320 \textdegree C with durations adjusted depending on particle size: 60 s for 100 nm, 180 s for 150 nm, and 300 s for 200 nm colloids.
For temperatures above 300  \textdegree C, a constant arsenic supply was maintained using tertiarybutylarsine (TBAs) with molar flow rates of 49 \textmu mol/min (100 nm and 200 nm samples) and 37 \textmu mol/min (150 nm samples). Trimethylgallium (TMGa) was used as the gallium precursor with a molar flow rate of 19 \textmu mol/min during both the pre-anneal growth step and subsequent NW growth.
Annealing was performed at 600 \textdegree C for 5 min under arsenic overpressure. NW growth was then carried out at 430 \textdegree C via the VLS mechanism.

Surface morphology and particle distributions were characterized using SEM (Hitachi S-4800) in secondary electron imaging mode. Images were taken at a 30\textdegree ~tilt angle at 10 kV acceleration voltage.

\begin{suppinfo}

Further details on the insufficient pure statistical modeling of the Au colloid deposition as a Poisson-like process as well as the general nature of the agglomeration increase for different colloid diameters are provided in the Supplementary Material.

\end{suppinfo}

\begin{acknowledgement}
    
The authors gratefully acknowledge the financial support received from the Carl-Zeiss-Stiftung for the “SustEntMat” project (funding code: P2023-02-008). Support by the Center of Micro- and Nanotechnologies (ZMN), a DFG-funded core facility (project number 233759584)  of the TU Ilmenau, is gratefully acknowledged.
The authors thank W. Prost for valuable discussions and constructive feedback on the manuscript and A. M{\"u}ller for
the experimental support.
GPT-5.3 (OpenAI) and DeepL Write were used solely for language editing, grammar improvement, and stylistic polishing of author-written text. All scientific content, data interpretation, and conclusions were reviewed and verified by the authors.
\end{acknowledgement}

\bibliography{Literature}

\includepdf[pages=-]{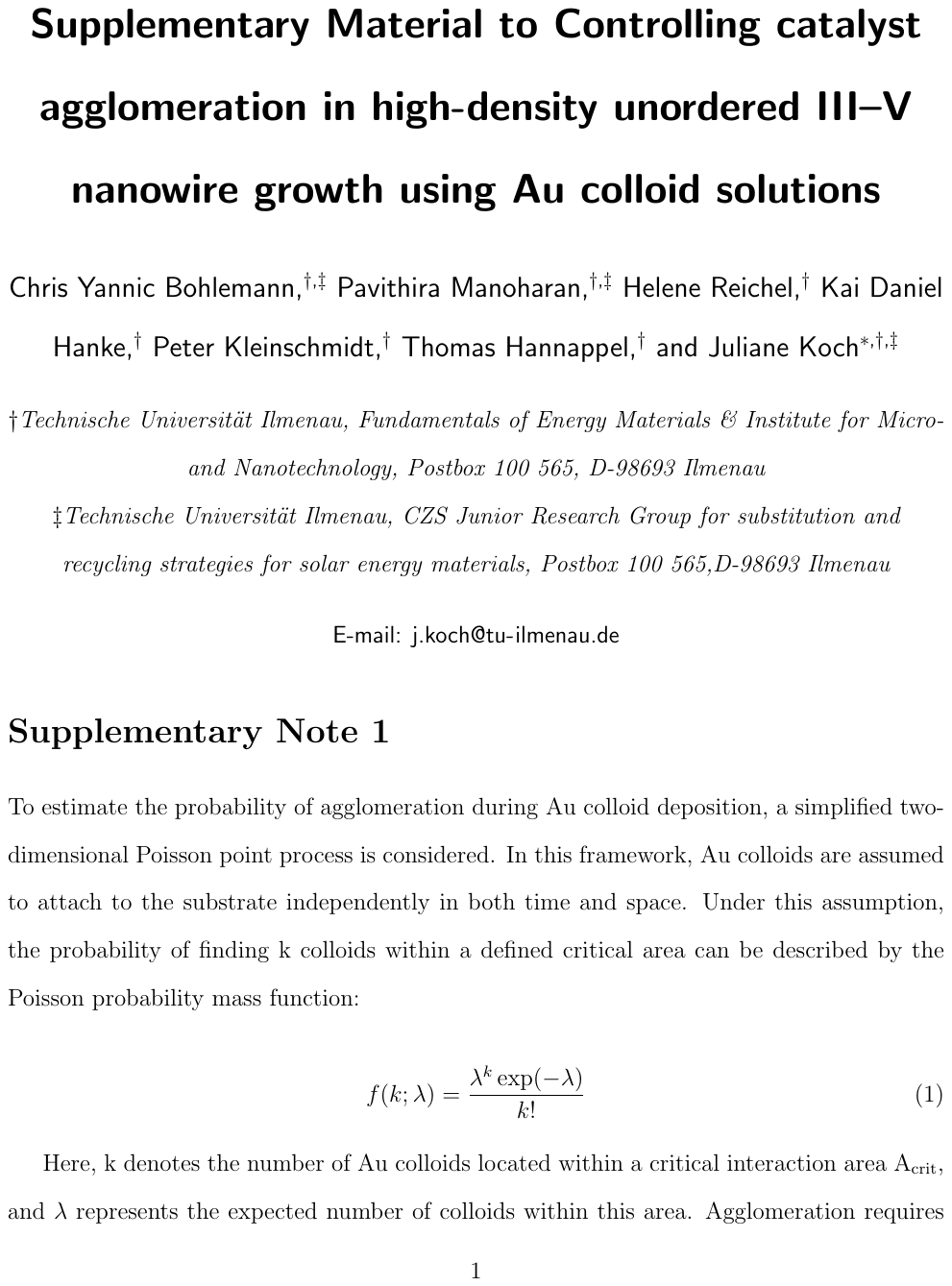}

\end{document}